\documentstyle[12pt,epsfig,amstex]{article}
\chardef\anciennecat=\catcode`\@
\catcode`\@=11
\renewcommand\section{\@startsection {section}{1}{\z@}%
                                   {-3.5ex \@plus -1ex \@minus -.2ex}%
                                   {2.3ex \@plus.2ex}%
                                   {\reset@font\large\bfseries}}
\catcode`\@=\anciennecat

\DeclareSymbolFont{AMSb}{U}{msb}{m}{n}
\DeclareMathSymbol{\C}{\mathalpha}{AMSb}{"43}
\DeclareMathSymbol{\R}{\mathalpha}{AMSb}{"52}
\DeclareMathSymbol{\Pom}{\mathalpha}{AMSb}{"50}
\DeclareMathSymbol{\T}{\mathalpha}{AMSb}{"54}
\newcommand{\xpom}{x_{\Pom}}
\newcommand{\dpom}{\Delta_{\Pom}}
\newcommand{\as}{\alpha_s}
\newcommand{\dg}[1]{\int\frac{d#1}{2i\pi}}

\begin{document}

\title{\bf Hard diffraction at HERA in the dipole model of BFKL dynamics}
\author{S. Munier, R. Peschanski {\it (SPhT)}\\
Ch. Royon {\it (DAPNIA/SPP)}\\
--- \\
 {\it Commissariat \`a l'Energie Atomique, Saclay}, \\
 {\it F-91191 Gif-sur-Yvette Cedex}\\
{\it France}
}
\maketitle

\begin{abstract} 
Using the QCD dipole picture of the hard BFKL pomeron, we derive the
general expressions of the elastic and inelastic components of the proton
diffractive structure functions.
We obtain a good 7 parameter fit (including a secondary reggeon contribution)
to data taken at HERA 
by the H1 and ZEUS
collaborations. The main characteristic features of the model 
in reproducing the data are discussed, namely the
effective pomeron intercept, the scaling violations and the beta 
dependence. A difference obtained in the separate H1 and ZEUS fits 
leads us to analyse directly the differences between both measurements.
Predictions on R, the ratio of longitudinal to transverse 
photon cross sections are performed and lead to 
very large values at high beta and large virtuality Q
which may lead to a discrimination between
models. 
\end{abstract}

\setcounter{equation}{0}
\setcounter{page}{1}


\noindent
\section{Introduction}

The HERA data for deep-inelastic scattering at high energy and 
high $Q^2$ contain a sizeable fraction of events
in which one observes a large rapidity gap in the forward region 
\cite{F2DH194,zeus}.
These events result from a colour-singlet exchange
between the diffractively dissociated virtual photon and the proton. 
Since this phenomenon is present even for high
virtuality of the photon,
it is called hard diffraction, at variance with soft diffractive phenomena.
Several theoretical formulations have been proposed \cite{tchanel}.
Among the most popular models, the one based on a pointlike structure of
the pomeron \cite{is} has been studied quantitatively 
using a non-perturbative input supplemented by a DGLAP evolution
\cite{F2DH194}.
In this formulation, it is assumed that the exchanged object, the pomeron, 
is a colour-singlet quasi-particle whose structure is probed in
deep-inelastic scattering.

There exists a different approach \cite{nz,bp} in which the cross sections 
are determined by the interaction between colour dipole states \cite{mueller}
describing the photon and the proton.
Indeed, it is well-known that the photon can be analyzed in terms of $q\bar{q}$ 
configurations \cite{bks} while
it has been shown \cite{mp} that the small-$x$ structure function of the proton
can be described by a collection of primordial dipoles with subsequent perturbative
QCD evolution. 
More specifically \cite{bp}, the combination of
the dipole description of perturbative QCD at high energy and the Good-Walker
mechanism \cite{gw} leads to a unified description of the proton total and
diffractive structure functions \cite{bpr}. However, a quantitative description
of the data following \cite{bp} was still lacking. 
The aim of the present paper is to provide a simple and
quantitative formulation fitting the experimental data.

In the dipole approach, two components are shown to contribute to the diffractive
structure function (see fig.1). First, an elastic component
(fig.1-a) corresponds to the elastic interaction of two dipole configurations.
It is expected to be
dominant in the finite $\beta$ region, i.e. for small relative masses 
of the diffractive system. 
Second, there is an inelastic component (fig.1-b) where the initial
photon dipole configuration is diffractively dissociated in multi-dipole states
by the target. This process is expected to be important at
small $\beta$ (large masses). Note that both components are obtained in ref.\cite{bp}
through perturbative QCD resummation corresponding to the BFKL approach \cite{bfkl}.
In these respects, it is based on a ``hard pomeron'' approach at variance with e.g.
a non perturbative pomeron input evolved using DGLAP evolution \cite{F2DH194}.

We present here fits of the published diffractive data \cite{F2DH194,zeus}
based on a simple with 7 free parameters and analytic parametrization for the 
diffractive structure functions in the QCD dipole model.  
Both fits are successful showing that this approach is a good candidate for an
understanding of hard diffraction. The main input of the calculation is
a formulation of the elastic QCD amplitude $T(r,\rho,b;\xpom)$
for the diffusion of a dipole of size $r$ on a dipole of size $\rho$ with rapidity
gap $\log 1/\xpom$  at fixed 
{\it impact parameter} $b$. This $b$-dependent input allows one to
factorize the different integrals involved in the perturbative calculation
of the diffractive components. The dipole model for the proton \cite{mp}
is used to provide predictions for the HERA data. 
We also include a phenomenological secondary Regge trajectory 
which is known to play a r\^ole in the limited domain of large mass and
small rapidity gap \cite{F2DH194}.

The outline of the paper is as follows: in section {\bf 2}, we introduce the dipole
formalism and in particular the input elastic dipole-dipole amplitude 
$T(r,\rho,b;\xpom)$ following the
requirements of the BFKL dynamics. Using this amplitude, we derive
the general expressions for the elastic (\ref{eqn:res2t}, \ref{eqn:res2l}), and
inelastic (\ref{eqn:res1tl}) diffractive components for both transverse and
longitudinal polarizations of the photon. 
In section {\bf 3} we obtain an analytical
form of the proton diffractive amplitudes used to fit the HERA data.
In section {\bf 4},  
the resulting fit is displayed. The simple form of the amplitudes makes 
transparent the discussion of the
model properties e.g. the relative contribution of the two components,
the effective exponent for the rapidity gap dependence, the origin of the
scaling violations, and the $\beta$ dependence.
Conclusions and outlook are given in the final section {\bf 5}.


\section{The QCD dipole formalism}

As mentionned in the introduction, the main input of our approach to
the structure functions is the interaction amplitude $T(r,\rho,b;\xpom)$ 
for a dipole of size $\rho$ with a dipole of size $r$, at impact
parameter $b$ where $\log 1/\xpom$ is the rapidity gap.
The knowledge of the amplitude as a function of $b$ is essential for the calculation of the 
two components of hard diffraction \cite{bp}. It is thus compulsory to start
with a correct BFKL amplitude in the whole impact parameter space which goes beyond
the commonly used expression for the forward elastic amplitude 
$T^{(forward)}=\int d^2b\; T(r,\rho,b;\xpom)$. 
For this sake, one uses the conformal invariance of the BFKL kernel in transverse
coordinates \cite{snw}. 
We shall propose such a $b$-dependent amplitude by asking for the two 
following requirements:\\
\noindent
{\bf (i)} The integral over impact parameter space boils down to the known exact expression 
for the forward amplitude $T^{(forward)}$ \cite{bfkl,snw}.\\
\noindent
{\bf (ii)} It has the correct high impact parameter approximation \cite{mueller}
including the modification by a scale factor which has been deduced from 
conformal invariance \cite{snw}.

Using the inverse-Mellin transform which appears in the solution of the BFKL equations
\cite{bfkl}, the amplitude reads:
\begin{equation}
T(r,\rho,b;\xpom)=\frac{1}{2}\frac{\rho^2 r^2}{b^4}
\dg{\gamma}(1-2\gamma)d(\gamma)\left(\frac{16 b^2}{r\rho}\right)^
{2\gamma}\xpom^{-\dpom(\gamma)}\ ,
\label{eqn:amplitude}
\end{equation}
where the factor 16 is the scale factor determined by conformal invariance.
\begin{equation}
\dpom(\gamma)=\frac{\as N_c}{\pi}\chi(\gamma);
\ \chi(\gamma)=2\psi(1)-\psi(\gamma)-\psi(1-\gamma)
\end{equation}
is the $\gamma$-dependent QCD pomeron intercept \cite{bfkl}, 
and
\begin{equation}
d(\gamma)=\frac{\as^2}{16\gamma^2(1-\gamma)^2}
\end{equation}
is the two-gluon exchange elementary dipole-dipole cross
section. 
As already mentioned, 
the amplitude meets the abovementionned requirements, since:\\
\noindent
{\bf (i)} the integration over $b$ leads to the known expression for
the forward amplitude (for $r>\rho$):
\begin{equation}
\int_{r/4}^{+\infty}
   \int_{r/4}^{+\infty}2\pi\;b\;db\;T(r,\rho,b;\xpom)= 8\pi\rho^2\dg{\gamma}d(\gamma)\left(
   \frac{r}{\rho}\right)^{2\gamma} \xpom^{-\dpom(\gamma)}\equiv T^{(forward)}\ .
\end{equation}
Note that the lower integration bound $r/4$ on $b$ is required
to recover the correct expression for the forward amplitude.
In the absence of an exact conformal invariant solution for all impact parameter
\cite{mnprsv}, this cutoff will play an important role in the determination
of the hard diffraction components.\\
\noindent
{\bf (ii)} A saddle point approximation of the integral over $\gamma$ in
formula (\ref{eqn:amplitude}) gives the approximate
expressions in the large $b$ approximation used in ref.\cite{bp,mueller}, 
up to the scale factor $16$.
This factor is strictly speaking not present in the large $b$ 
approximation, but appears when one takes into account the global
conformal invariance of the BFKL kernel \cite{snw}. This scale factor
leads to a much more central impact parameter distribution than the 
approximate expressions used
in ref.\cite{bp,mueller}, and plays a major r\^ole in diffraction, as we 
shall see in the following.

Using formula (\ref{eqn:amplitude}) as an input leads to an analytic formulation
of the diffractive amplitudes. In this section, we shall first consider 
hard diffraction on
a dipole of given size $r$. It will be shown in the next section how, starting from
this case, it is possible to derive phenomenological expressions for hard diffraction
on a proton.
One has the following identities for the diffractive structure functions 
\cite{bp}:
\begin{equation}
F_{T,L}^D(Q^2,\xpom,M^2)=\frac{Q^4}{4\pi^2\alpha_{em}}\frac{1}{\beta\xpom}
\frac{d\sigma_{T,L}}{dM^2}
 =\frac{Q^2}{4\pi^2\alpha_{em}}\frac{1}{\xpom}\beta\frac{d\sigma_{T,L}}{d\beta}\ ,
\label{eqn:sectionf}
\end{equation}
where $d\sigma_T$ (resp. $d\sigma_L$) is the differential cross 
section for the scattering of a virtual photon with transverse 
(resp. longitudinal) polarization, on a dipole.

\vskip 1em


\noindent{\bf a)} {\it Elastic component}

\noindent
Using the QCD dipole model for the elastic component (see fig. 1.a), one
writes
\begin{equation}
\frac{d\sigma^{(el)}_{T,L}}{dM^2}=\frac{N_c\alpha_{em}e^2}{2\pi}Q^2
  \int_{r/4}^\infty d^2b \int_0^1dz f_{T,L}(z)\left|\T_{T,L}(b;\xpom)
\right|^2
\label{eqn:dsdm}
\end{equation}
where
\begin{equation}
\T_{T,L}(b;\xpom)=\int_0^{4b} d\rho\;\rho\;T(r,\rho,b;\xpom)
   K_{1,0}(\hat{Q}\rho)J_{1,0}(\hat{M}\rho)
\label{eqn:deft}
\end{equation}
and
\begin{align}
f_T(z)&\equiv z^2(1-z)^2(z^2+(1-z)^2)\nonumber\\
f_L(z)&\equiv 4z^3(1-z)^3\ .
\label{eqn:f}
\end{align}
The notation ``~$\hat{X}$~'' is defined by $\hat{X}\!=\!(z(1-z))^{1/2}X$.

Following ref.\cite{bp}, the formulae (\ref{eqn:dsdm}--\ref{eqn:f})
describe the interaction of a $q\bar{q}$ configuration of transverse size $\rho$ of the
virtual photon with a dipole of size $r$.
The input elastic amplitude is taken from 
(\ref{eqn:amplitude}). 
The photon wave function in terms of $q\bar{q}$ configurations 
is projected on its transverse ($T$) and longitudinal ($L$) 
polarization states which differ only by the factors $f_{T,L}(z)$
and by the order of the Bessel functions \cite{grad} involved: $K_1$, 
$J_1$ are associated to the transverse component, and $K_0$, $J_0$ 
to the longitudinal one. $z$ (resp. $(1-z)$) is the longitudinal momentum fraction
carried by the antiquark (resp. quark). 
Note that interference effects are present \cite{bp} which explain the product
of Bessel functions of different kinds ($K$ and $J$).
The cutoff in impact parameter space appears
twice in formulae (\ref{eqn:dsdm}) and in (\ref{eqn:deft}), 
first by the upper bound $4b$ on $\rho$ and second by a lower bound 
$r/4$ on the impact
parameter $b$. In practice, we will release the first bound, which will
greatly simplify the calculations and allow to find an analytical form
for the diffractive amplitude. This approximation is expected to change only the
normalisations.

Inserting expression (\ref{eqn:amplitude}) for the dipole-dipole amplitude 
in eq. (\ref{eqn:dsdm}, \ref{eqn:deft}), one obtains the results  
for $F_{T,L}^D$ (\ref{eqn:sectionf})
involving integrals over the impact parameter $b$, the momentum fraction $z$
and the $q \bar{q}$-pair size $\rho$.
Interestingly enough, these integrations can be factorized and exactly
computed using in particular the change of variable 
$\rho\rightarrow\hat{\rho}=(z(1-z))^{1/2}\rho$. 
After a tedious but straightforward calculation using well-known identities
\cite{grad}, the successive integrations give
\begin{multline}
F_T^{D(el)}=\frac{16^{3}}{2\xpom}\frac{N_ce^2}{\pi^2}\beta^2(1-\beta)
  \left(\frac{Q_0}{Q}\right)^2
  \int{\underset{j=1,2}\prod}\left[\frac{d\gamma_j}{2i\pi}
  \left(\frac{Q}{2Q_0\sqrt{\beta}}\right)^
{2\gamma_j}\xpom^{-\dpom(\gamma_j)}\right]\\
 \times{\underset{j=1,2}\prod}\left\{(1\!-\!2\gamma_j)d(\gamma_j)
 \Gamma(2\!-\!\gamma_j)\Gamma(3\!-\!\gamma_j)
 {\sideset{_2}{_1}F}(2\!-\!\gamma_j,-1\!+\!\gamma_j;2;1\!-\!\beta)\right\}\\
  \times\frac{\gamma_1\!+\!\gamma_2}{(\gamma_1\!+\!\gamma_2\!-\!1)
  (3\!-\!2(\gamma_1\!+\!\gamma_2))}
    B(\gamma_1\!+\!\gamma_2,1/2)\ ,
\label{eqn:res2t}
\end{multline}

\begin{multline}
F_L^{D(el)}=\frac{16^{3}}{\xpom}\frac{N_ce^2}{\pi^2}\beta^3
  \left(\frac{Q_0}{Q}\right)^2
  \int{\underset{j=1,2}\prod}\left[\frac{d\gamma_j}{2i\pi}
  \left(\frac{Q}{2Q_0\sqrt{\beta}}\right)^
{2\gamma_j}\xpom^{-\dpom(\gamma_j)}\right]\\
 \times{\underset{j=1,2}\prod}\left\{(1\!-\!2\gamma_j)d(\gamma_j)
 \Gamma^2(2\!-\!\gamma_j)
 \;{\sideset{_2}{_1}F}(2\!-\!\gamma_j,-1\!+\!\gamma_j;1;1\!-\!\beta)\right\}\\
  \times\frac{1}{3\!-\!2(\gamma_1\!+\!\gamma_2)}
  B(\gamma_1\!+\!\gamma_2,1/2)\ ,
\label{eqn:res2l}
\end{multline}
where we have introduced the scale $Q_0\equiv 2/r$.
Note that these expressions involve two coupled inverse-Mellin transforms
in $\gamma_1$ and $\gamma_2$, each corresponding to one elastic dipole interaction, 
(see fig.1-a). 

\vskip 1em


\noindent{\bf b)} 
{\it Inelastic component}

\noindent
As shown in fig. 1.b, the inelastic component
stems from the following process: the initial $q\bar{q}$ state of the virtual
photon develops a set of colour dipoles through cascading \cite{mueller}
and the diffractive component is due to the interactions of two of the produced
dipoles with a target dipole, each of these being described by the amplitude 
(\ref{eqn:amplitude}).
The cross section can be expressed as \cite{bp}:
\begin{equation}
\beta\frac{d\sigma^{(in)}_{T,L}}{d\beta}=16^{3}\alpha_{em}e^2\as^5 N_c^2\frac{1}{Q^2}
\int_{r/4}^{\infty}
\frac{d^2b}{b^2}\left<{\varrho}^2\right>^2
\dg{\gamma}\left(\frac{b\;Q}{2}\right)^{2\gamma}
(1-\gamma)^3 H_{T,L}(\gamma)\beta^{-\dpom(\gamma)}\ ,
\label{eqn:comp1}
\end{equation}
where
\begin{align}
\left<\varrho^2\right>&=\int_0^{4b}\frac{d\varrho}{\varrho}
  \varrho^2\frac{T(r,\varrho,b;\xpom)}{4\pi\as^2\varrho^2}
\label{eqn:rho2}
\end{align}
and
\begin{align}
H_T(\gamma)&=\frac{V(\gamma)}{16\gamma^2(1\!-\!\gamma)^4}
  \frac{\Gamma(3\!-\!\gamma)\Gamma^3(2\!-\!\gamma)\Gamma(2\!+\!\gamma)
        \Gamma(1\!+\!\gamma)}
       {\Gamma(4\!-\!2\gamma)\Gamma(2\!+\!2\gamma)}\nonumber\\
H_L(\gamma)&=\frac{V(\gamma)}{8\gamma(1\!-\!\gamma)^3(1\!+\!\gamma)(2\!-\!\gamma)}
  \frac{\Gamma(3\!-\!\gamma)\Gamma^3(2\!-\!\gamma)\Gamma(2\!+\!\gamma)
        \Gamma(1\!+\!\gamma)}
       {\Gamma(4\!-\!2\gamma)\Gamma(2\!+\!2\gamma)}
\label{eqn:h}
\end{align}
with
\begin{equation}
V(\gamma)={\sideset{_3}{_2}F}(1\!-\!\gamma,1\!-\!\gamma;1,3/2;1).
\label{eqn:v}
\end{equation}

The dipole-dipole interaction is represented by formula (\ref{eqn:rho2})
which can be interpreted as the average squared transverse sizes
of the dipoles seen at the impact parameter $b$. Indeed, the quantity 
$T(r,\varrho,b;\xpom)/4\pi\as^2\varrho^2$ is nothing else than the density of
dipoles of size $\varrho$ emitted from a dipole of size $r$ at impact 
parameter $b$ \cite{mueller,snw}.
The inverse Mellin-transform and its integrand factor $V(\gamma)$ (\ref{eqn:v})
comes from the calculation of the $1\rightarrow 2$ dipole branching at the virtual 
photon vertex \cite{bp}.
The different prefactors appearing in  $H_T$ and $H_L$ in formula (\ref{eqn:h})
come from the transverse and longitudinal wave-functions of the virtual photon, 
respectively.

Inserting the expression (\ref{eqn:rho2}) into eq. (\ref{eqn:comp1})
and integrating over the impact parameter $b$, one gets:
\begin{multline}
F_{T,L}^{D(in)}=\frac{1}{\xpom}\frac{16^{5}}{\pi^3}
e^2\as N_c^2
  \int{\underset{j=1,2}\prod}\left\{\frac{d\gamma_j}{2i\pi}
  \frac{1\!-\!2\gamma_j}{2\!-\!2\gamma_j}
   d(\gamma_j)\xpom^{-\dpom(\gamma_j)}\right\}\\ 
\times\dg{\gamma}\left(\frac{Q}{4Q_0}\right)^{2\gamma}\beta^{-\dpom(\gamma)}
  (1\!-\!\gamma)^3H_{T,L}(\gamma)
  \frac{1}{2\!-\!\gamma\!-\!\gamma_1\!-\!\gamma_2}\ .
\label{eqn:res1tl}
\end{multline}
The formulae (\ref{eqn:res2t}, \ref{eqn:res2l}, \ref{eqn:res1tl}) condense
in an analytical form the perturbative QCD predictions for the diffractive 
structure functions of a dipole of size $2/Q_0$.


\section{Proton diffractive structure functions in the dipole model}

In the previous section, we have obtained the expressions for
the diffractive structure functions, when the target is a single dipole
of size $r=2/Q_0$. Our aim in this section is to use these formulae
to obtain the diffractive structure functions of the proton.
We have thus to implement a model for the proton which could describe
it as a collection of primordial dipoles.

\vskip 1em

\noindent{\bf a)} {\it Elastic component}\\
\noindent
Applying the steepest-descent method,
the saddle-point is determined by the usual BFKL-type integrand (between the
square brackets in (\ref{eqn:res2t}, \ref{eqn:res2l})). It reads:
\begin{equation}
\gamma_{sp}=\frac{1}{2}\left(1-a(\xpom)\log\frac{Q}{2Q_0\sqrt{\beta}}\right)\ .
\end{equation}
where $a(x)=\pi/(7\zeta(3)\as N_c \log 1/x)$ is the $k_\bot$-diffusion 
coefficient coming from the BFKL evolution \cite{bfkl,bartels}.
We have to use a special treatment for the pole 
at $\gamma_1+\gamma_2=1$ in formula (\ref{eqn:res2t}), since 
it contributes for the asymptotic
saddle points at $\gamma_1\!=\!\gamma_2\!=\!1/2$ when $\xpom\rightarrow 0$. 
All the other singularities are not relevant, since they are far from the saddle points.

Noting that in the kinematical domain of interest,
$1-2\gamma_{sp}=a(\xpom)\log Q/2Q_0\sqrt{\beta} \ll 1$,
and working at lowest order, the results are:
\begin{multline}
F_T^{D(el)}=12\frac{N_ce^2\as^4}{\pi}
\xpom^{-2\alpha_\Pom+1}\;a^3(\xpom)
\log^3\frac{Q}{2Q_0\sqrt{\beta}}\;e^{-a(\xpom)\log^2\frac{Q}{2Q_0\sqrt{\beta}}}\\
\times\beta(1\!-\!\beta)\;
\left[{\sideset{_2}{_1}F}\left(-\frac{1}{2},\frac{3}{2};2;1-\beta\right)
\right]^2\ ,
\label{ftdqef}
\end{multline}
\begin{multline}
F_L^{D(el)}=16\frac{N_ce^2\as^4}{\pi}\;
\xpom^{-2\alpha_\Pom+1}\; a^3(\xpom)\log^2\frac{Q}{2Q_0\sqrt{\beta}}
\;e^{-a(\xpom)\log^2\frac{Q}{2Q_0\sqrt{\beta}}}\\
\times\beta^2\;
\left[{\sideset{_2}{_1}F}\left(-\frac{1}{2},\frac{3}{2};1;1-\beta\right)\right]^2\ ,
\label{fldqef}
\end{multline}
with
\begin{equation}
\alpha_\Pom\equiv 1+\dpom\left(\frac{1}{2}\right)=1+\frac{\as N_c}{\pi}4\log 2\ .
\label{defap}
\end{equation}
The pole at $\gamma_1\!\!+\!\!\gamma_2\!\!=\!\!1$ in $F_T^{D(el)}$ results in an extra 
$\log Q/2Q_0\sqrt{\beta}$ which
may be large.
\par 
The formulae (\ref{ftdqef}, \ref{fldqef}) give
an explicit expression for the elastic components
of diffraction in the QCD dipole model.
The $\xpom$-dependent prefactor $\xpom^{-2\alpha_\Pom\!+\!1}a^3(\xpom)$ plays the role of the
(hard) pomeron flux factor already discussed in references \cite{bp,bpr}. Revealing new 
features of the model, the present explicit
calculation shows that the relevant scale factor responsible for the scaling violations
is $2Q_0\sqrt{\beta}$ where $Q_0$ is the scale of the proton-dipole amplitude.
The $\beta$-dependence is found to be non polynomial.
\vskip 1em
\noindent{\bf b)} {\it Inelastic component}

\noindent
Using again the steepest-descent method for evaluating the three integrals in $\gamma$, 
$\gamma_1$ and $\gamma_2$ of formula (\ref{eqn:res1tl}), we find saddle-points at
\begin{equation}
\gamma_{sp}=\frac{1}{2}(1-a(\beta)\log \frac{Q}{4Q_0})\ ,
\ \gamma_{j,sp}=\frac{1}{2}(1-a(\xpom))\ .
\end{equation}
Note that there is no coupling of the integrals at the saddle points;
however, one should take into account the prefactor zeros at $\gamma_j=1/2$, see
eq. (\ref{eqn:res1tl}).
All in all, the result reads:
\begin{equation}
F_{T,L}^{D(in)}=2^9\sqrt{\frac{2}{\pi}} 
H_{T,L}\left(\frac{1}{2}\right)\frac{N_ce^2\as^5}{\pi^4}
\;\xpom^{-2\alpha_\Pom+1}a^3(\xpom)\frac{Q}{Q_0} 
e^{-\frac{a(\beta)}{2}\log^2\frac{Q}{4Q_0}}
 a^{\frac{1}{2}}(\beta)\beta^{-\dpom}\ .
\label{fdinf}
\end{equation}
Comparing our expression (\ref{fdinf}) with the original result (formula (11)
in the second paper of ref.\cite{bp}), we find identical results but for the scale
of the transverse momentum in the BFKL evolution: $4Q_0$ 
instead of $Q_0$. This change comes from the scale factor
determined by conformal invariance (see 2.1)
and was not considered in the previous approximation \cite{bp}. 
This modified scale factor has an important phenomenological effect.

In the QCD dipole model, there is a tight connection between the total and
diffractive structure functions. In refs.\cite{mp,bpr}, 
the QCD dipole model has
been successfully applied to the proton {\it total} structure function in the
small-$x$ kinematical domain. A model for the primordial dipole
configurations in the proton has been introduced through a distribution with
average size and the non perturbative features of the proton
target manifest themselves only through this scale and the global normalization 
constant related to the density of primordial dipoles (for a more complete discussion,
see ref.\cite{mp}).
We thus use this representation of the nucleon target in formulae 
(\ref{ftdqef}, \ref{fldqef}, \ref{fdinf}) to derive the model for the
proton diffractive structure functions.
In this context, the scale parameter is reinterpreted as $Q_0=2/\!\!<\!\!r\!\!>$ 
and we introduce
the arbitrary normalizations $N_T^{(el)}$, $N_L^{(el)}$, $N^{(in)}$
for $F_{T,L}^{D(el)}$ and $F_{T,L}^{D(in)}$ multiplying formulae 
(\ref{ftdqef}, \ref{fldqef}, \ref{fdinf}) respectively.
These unknown normalizations reflect the non-perturbative primordial dipole
configurations contributing to the various proton structure functions.

With this model, we are ready to write a full parametrization adequate for the 
description of the data.
The free parameters of the dipole model are $\alpha_\Pom$, which is related to
the fixed coupling constant $\as$ in the BFKL scheme at leading
order (see eq. (\ref{defap})), $Q_0$, corresponding to a non-perturbative scale
for the proton, and the three normalizations $N_T^{(el)}$, $N_L^{(el)}$, $N^{(in)}$.
As is now well-known, a secondary
trajectory based on reggeon exchange is added in order to take into account the
large-mass and small rapidity gap domain.
Reggeon exchange can here be simply parametrized in the following way:
\begin{equation}
F_2^{D(R)} (x_P, \beta, Q^2) =
 f^\R (x_\Pom) F_2^\R (\beta, Q^2)\ ,
\label{regge}
\end{equation}
where the reggeon flux $f^\R (x_\Pom)$ is assumed to follow a Regge behaviour with a linear
trajectory $\alpha_{\R}(t)=\alpha_{\R}(0)+\alpha^\prime_{\R}\;t\;$:
\begin{equation}
f^{\R}(x_{\Pom})= \int^{t_{min}}_{t_{cut}} dt\;\frac{e^{B^{\R}\; t}}
{x_{\Pom}^{2 \alpha_{\R}(t) -1}}\ ,
\end{equation}
where $|t_{min}|$ is the minimum kinematically allowed value of $|t|$
and $t_{cut}=-1$ GeV$^2$ is the limit of the measurement. The values
of $B^{\R}$ and $\alpha^\prime_{\R}$ are fixed with data from hadron-hadron
collisions \cite{F2DH194}. The reggeon structure function is assumed 
to be the pion structure function \cite{owens}. The free parameters for 
this component are the reggeon normalisation $N^\R$ and 
exponent $\alpha_{\R}$.
The final parametrization used for the fit is the sum of the
three contributions detailed above (\ref{ftdqef}, \ref{fldqef}, \ref{fdinf},
\ref{regge}), namely:
\begin{eqnarray}
F_{2}^{D(3)}= \frac{1}{N_C e^2} \left( N^{in} (F^{D(in)}_T +F^{D(in)}_L) 
 +  N^{el}_{T}   F^{D(el)}_T
+ N^{el}_{L} 
 F^{D(el)}_L + N_R  F_2^{\R} \right).
\label{eqnfinal}
\end{eqnarray}
Only one parameter ($N^{in}$) is used for the normalisation of the
inelastic component since $F^{D(in)}_L / F^{D(in)}_T = 2/9$.


\section{Phenomenology}
\subsection{Fits to the H1 and ZEUS data}
A fit to the recently published H1 \cite{F2DH194} and ZEUS
\cite{zeus} diffractive structure 
function data is performed separately. The result of both fits is
shown in Figure 2 for H1 and Figure 3 for ZEUS, and their parameters are
given in Table 1. Let us comment the interesting features of the fits. 
The fit to the H1 data leads
to a very good $\chi^2$ (1.17 per degree of freedom with statistical errors
only). 
The value of $\alpha_\R$ is consistent 
with the usual values found for secondary reggeon contributions if interference effects
are taken into account \cite{F2DH194}. The value of $\alpha_\Pom$
is found to be consistent with the expected intercept for a hard BFKL pomeron
\cite{bfkl}. This intercept is higher than the value obtained from the
fit to the structure function $F_2$ \cite{mp}.  $Q_0$ is a typical non perturbative scale for the
proton and very close to the value obtained in the proton structure function
fit. It should be noted that the scale $Q_0$ appears in a quite non trivial
way as the virtuality in the inelastic component ($Q/4Q_0$), and in the
elastic one ($Q/2 \sqrt{\beta} Q_0$). The validity of these results can
be checked by starting with two different values of $Q_0$ for each component and
the result of the fit leads to same values within errors, with the same
$\chi^2$. Furthermore, imposing a constant scale for the elastic component, i.e. $Q/2 \sqrt{\beta} Q_0 \to Q/Q_0,$
leads to a very bad quality fit. The other features
of this description will be discussed further on.
\par
The fit to the ZEUS data leads to a worse $\chi^2$ ($\chi^2 /dof=
1.95$ with statistical errors only). 
The result is shown in Figure 3 in continuous line, and by comparison,
the dashed curve corresponds to the parameters obtained in the H1 fit. 
It can be noticed that a difference is clearly seen both in the
low $Q^2,$  low $\beta$ and  the
high $Q^2,$  high $\beta$ bins, in opposite way, the H1 fit going from below to above the Zeus fit. 
\par
In order to investigate the origin of these differences, a direct comparison
between ZEUS and H1 data has been performed \cite{zeuthen}. The H1 data have
been interpolated to the ZEUS closest bins in $\beta$ and $Q^{2}$ using the
dipole model fit. This interpolation is weakly sensitive to the model used
as the interpolation in the kinematical variables is very small. It was 
checked that the use of the model 
by Bartels et al. \cite{bal} gives a similar result. The result of this comparison is
displayed in Figure 4. The striking feature is that the main difference
between both fits noticed in Figure 3 comes from the region where the data are most
different. The differences occur both at small and high $\beta$ (i.e. small and
high masses) and the solution of this puzzle cannot be due only to the 
difference in selecting diffractive events 
- either the rapidity gap selection \cite{F2DH194} or the $M_X$
substraction method \cite{zeus} - in both experiments. In fact,
only high-mass events could be affected by these different selections. 
A global shift of ZEUS data of about 15\% would be possible as they do
not correct for proton dissociation but the difference between H1
and ZEUS data is not only due to normalisation effects. Indeed, it is
clear in Figure 4 that, depending on the values of $Q^2$ and $\beta$,
H1 points lie either higher or lower than ZEUS data.
The analysis of new accumulated data in this kinematical region would help
solving the experimental puzzle and would be of great interest.

\subsection{Effective pomeron intercept}
Let us now study the dependence on $\xpom$ which is directly related 
to the rapidity gap dependence. One can define an effective pomeron 
intercept in the following way:
\begin{eqnarray}
\alpha_{\Pom}^{eff} = \frac{1}{2} \left( \frac{d ln F_2^{D}}{d ln 1/ x_{\Pom}}
+ 1 \right)
\end{eqnarray}
where the $t$ dependence is integrated out (the data are mainly at $t
\sim 0$). This effective exponent can also be determinated for the
inelastic and elastic components separately as a function of
$\xpom$ (Figure 5) or $Q^2$ (Figure 6). 

In Figure 5, The effective intercepts of both
components and their sums have been compared to the soft pomeron intercept \cite{dola}
(dashed line)  and to the bare hard BFKL pomeron obtained in the 
fit ($\alpha_{\Pom}=1.395,$ cf. Table 1). The range of obtained values sits essentially 
between these two limits except in the large $\xpom$ region ($\xpom \ge 
10^{-2}$). It is clearly not consistent with the soft pomeron value
(1.08). It is as well much lower than the bare pomeron intercept
\cite{bp,bpr}. This can be explained by the large logarithmic
corrections induced by the $a^3(x_\Pom)$ term, proportional to
$\log^3(1/x_\Pom)$, present in both diffractive components (see formulae 
\ref{ftdqef}, \ref{fldqef}, \ref{fdinf}). The effect of this logarithmic
term induces also an $\xpom$ dependence of the intercept. Moreover, in Figure 5,
it can be seen that the $\xpom$ dependence of the intercept is different 
between the elastic and the inelastic components. This induces a breaking of
factorisation directly for the diffractive components of this model, which 
comes in addition to the known factorisation breaking due to secondary 
trajectories. In addition, the elastic component itself does not factorise,
as it can be seen in Figure 5 for different $Q^2$ values. This comes from
the mixing of the $\xpom$ and $Q^2$ dependence in formula \ref{ftdqef}, 
\ref{fldqef}.
\par
As seen in Figure 6, the $Q^2$ dependence of the pomeron intercept (taken
for instance at $x_\Pom$=0.001 where the reggeon contribution is negligible)
is weak. It is in good agreement with the H1 determination 
($\alpha_\Pom= 1.204 \pm .02$) indicated in Figure 6.
It is thus intermediate between the soft and hard pomeron lines but inconsistent
with both of them. In our model, this softening of the bare pomeron comes from the
large logarithmic corrections which cannot be neglected in perturbative
calculations as was mentioned in the last paragraph.

\subsection{Scaling violations}
One striking feature of the diffractive proton structure functions was 
the $Q^2$ dependance at fixed $x_{\Pom}$ as a function of $\beta$ as was
pointed out experimentally by the H1 collaboration \cite{F2DH194} 
(see Figure 7) and
confirmed at lower $Q^2$
\cite{F2DH195}. The structure functions are increasing with $Q^2$ even at
very high $\beta$ (see Figure 7) at variance with the behaviour of the total proton
structure function as a function of $x$ .
In the QCD dipole model, this experimental feature is described by a non
trivial interplay between the two diffractive components. In Figure 7,
the dipole fit is compared with the H1 result showing the contribution
of each component: at small $\beta$, the inelastic component dominated and
vary quasi linearly in $\log Q^2$, and at high $\beta$, this component
is depressed similarly to the total structure function, but is 
progressively substituted
by the elastic component. Note that the enhancement present in the data at
high $\beta$ and low $Q^2$ which is probably due to vector meson production
is partially reproduced by the model. It is expected that including the
specific vector meson contributions in addition to the elastic $q \bar{q}$
component will even improve the result of the model in this kinematical
range. It is striking that it is possible to describe the observed scaling
violations in a very different framework as the one given by the DGLAP
evolution as was performed by the H1 collaboration \cite{F2DH194}.

\subsection{$\beta-$dependence}
In Figure 8 is displayed the $\beta$ dependence of the proton diffractive
structure function for a fixed value of $x_{\Pom}=0.003$ for different
values of $Q^2$. The $\beta$ dependence is quite weak in all $Q^2$ bins
and correctly reproduced by the interplay of the two components of the QCD
dipole model. While at low $Q^2$, this effect is essentially due to the 
inelastic component, at high $Q^2$ the interplay between both components
is required to describe the observed $\beta-$dependence.
\par
The interplay of the different components 
can be analysed in more details in Figure 9 where the elastic, inelastic,
reggeon components and their sum are displayed for three different values
of $x_\Pom$ and four different values of $Q^2$ as a function of $\beta$.
The reggeon component is only important at high value of $x_\Pom$ as 
expected and dominates at low $\beta$. It disappears at smaller $x_\Pom$.
At low $Q^2$, the inelastic component dominates in almost the full $\beta$
range, while at higher $Q^2$, it is only important at low $\beta$ and
replaced at low masses by the inelastic component. 

\subsection{Longitudinal contribution to the diffractive structure
function}
In Figure 10 are displayed as a function of $\beta$ in different $Q^2$,
 $x_\Pom$ bins the longitudinal and transverse components of the
proton diffractive structure functions. Note that the $x_\Pom$ values
are chosen a bit smaller than in the previous figure to depress the
reggeon contribution. Note also that we did not separate the longitudinal
and the transverse contributions for the inelastic component as they
are directly proportional by a factor $ H_L(1/2)/H_T(1/2)=2/9$ (see
formula \ref{fdinf}). In addition to the already mentioned dominance of the
inelastic component at small $Q^2$, the longitudinal elastic component is
found to be high at high $\beta$ and crosses over the transverse component
near $\beta \sim 0.8$. It is thus expected to obtain high values of the
ratio $R$ of the longitudinal to the transverse components at high values 
of $\beta$. The result of our fit prediction is displayed in Figure 11.
We note that the $R$ ratio remains small ($\sim 0.2$) in almost the full
kinematical plane except notably at high $\beta$ where it may reach high values
such as 2. Note that this value is in the range of the measured $R$
ratio with vector meson production \cite{tchanel}. A measurement of $R$
in diffraction would thus be of great interest and would be a good test
of the model. The question arises whether such high values of $R$ would not
modify the $F_2^D$ measurement itself as it was assumed to be 0
\cite{F2DH194}. We have checked that the parameters and the quality of
the fit are not much influenced if only data points with $y<0.3$, where
the influence of $R$ is small, are
taken. 
\par 
It is instructive to notice that another model of diffraction based on
selecting $q \bar{q}$ and $q \bar{q} g$ components of the photon
\cite{bal} also leads to a large contribution of 
the longitudinal $q \bar{q}$ contribution at high $\beta$. However,
a distinct feature of the dipole model is the above mentioned difference
between the high and low $Q^2$ behaviour where at low $Q^2$, the
inelastic component dominates and induces a small value of $R$. We thus
expect thet a measurement of $R$ in diffraction at low $Q^2$ would be a 
way to distinguish both models.

\section{Conclusion and outlook}

Let us summarize the results we have obtained.
First of all, the colour dipole model has provided us with a interesting theoretical and phenomenological 
framework to study hard diffraction. On thetheoretical ground, 
it enables to interpret hard diffraction in terms of BFKL dynamics of (resummed)
perturbative QCD, which allows one to make theoretical estimates. On the other hand, the success of the phenomenological fit using analytical amplitudes allow a discussion of the various interesting and intringuing aspects of the data.

The model predicts two  contributions: an elastic one for
which the final state consists in the hadronization of a quark-antiquark
pair, and an inelastic one, including soft gluon radiation reproduced in terms of a cascading process involving colour dipoles. Both contribute in 
quite different kinematical
domains since they are characterized by different mass  distributions (different $\beta$) and for different photon  polarization.
In this paper, we have obtained a  suitable parametrization and a  satisfactory  fit to the
H1 data along these lines. The fit to Zeus data is less satisfactory and with different parameter values. By a model-independent analysis of both data, we could identify the reason of the discrepancy in a genuine difference between data sets when interpolating in the same bins. This point deserves a detailed experimental study in the near future, since it may be of importance to doiscriminate between models. in particular the confirmation of the trend of present H1 data would be in favor of the dipole model.

On the theoretical ground, we have computed amplitudes obeying some of the requirements
of BFKL dynamics for diffractive amplitudes (on a dipole). However, exact calculations of these rather complicated amplitudes are still under way \cite{exact} and we hope in the near future obtain definite predictions with all requirements of BFKL dynamics. the evaluation of next-leading BFKL effects would
also be welcome since they already play a r\^ole for the total structure functions.

Finally, as emphasized in ref.\cite{bpr}
the colour dipole model formalism calls for a unified description of the diffractive and  total
deep-inelastic scattering events, e.g. including events with no rapidity gap.
We showed that within the precision of the current data, there are quite a few 
indications (similar scale $Q_0,$ softening of the gard Pomeron by logarithmic facors in diffraction,etc...) of such a common theoretical ground. 
However, further tests of the model are deserving. The first one would be a confrontation
of the predicted $R$ ratio with the data if available: indeed, the various models
should predict quite different contributions from the two polarization states
of the photon. Other useful tests concern the
final states. For instance one can compute the 
predictions for diffractive vector meson production and confront them to the recent data. Such tests might help distinguish between
the few different models for hard diffraction which are able to describe the data. definitely, the QCD dipole model of BFKL dynamics is one of them.

\section{Acknowledgements}
A fruitful collaboration with Andrzej Bialas and Henri Navelet which has much
inspired the present paper is acknowledged. We also thank them for very useful
physical and technical remarks and suggestions during the elaboration of the paper.
\eject


\newpage

\noindent
{\bf TABLE CAPTION}

\vspace{1cm}
\noindent
{\bf Table I} 

\vspace{0.5cm}
\noindent
{\it Parameters obtained for the $F_2^D$ fit.}

\vspace{0.5cm}
The fit has been performed with statistical errors only.
The first error quoted is the statistical one and the second the
systematic one.

\vskip 3cm

\begin{center}
\begin{tabular}{|c|c|c|} \hline
 & H1  & ZEUS\\
\hline\hline
$\alpha_{P}$ & 1.395 $\pm$ 0.005 $\pm$ 0.003 & 1.327 $\pm$ 0.001 $\pm$ 0.016\\
$\alpha_{R}$ & 0.682 $\pm$ 0.046 $\pm$ 0.049 & --- \\
$Q_0$ & 0.428 $\pm$ 0.011 $\pm$ 0.001 & 0.241 $\pm$ 0.002 $\pm$ 0.014\\ 
\hline
$N^{in}$ & 0.00244 $\pm$ 0.00028 $\pm$ 0.00033 & 0.00374 $\pm$ 0.00015 $\pm$ 0.00076\\
$N^{el}_T$ & 40.0 $\pm$ 1.6 $\pm$ 3.6 & 126.9 $\pm$ 0.5 $\pm$ 37.6 \\
$N^{el}_L$ & 12.8 $\pm$ 1.2 $\pm$ 1.1 & 24.43 $\pm$ 0.51 $\pm$ 8.12 \\
$N_{R}$ & 7.44 $\pm$ 1.84 $\pm$ 3.50  & 0 $\pm$ 58 $\pm$ 0 \\ \hline
$\chi^2$ & 255.4\ \mbox{for}\ 226\ \mbox{pts} & 89.6\ \mbox{for}\ 53\ \mbox{pts}\\
         & 1.17/\mbox{d.o.f.} & 1.95/\mbox{d.o.f.}\\ \hline
\end{tabular}
\end{center}
\vskip 1cm
\begin{center}
{\large\bf Table I}
\end{center}

\clearpage

\noindent
{\bf FIGURE CAPTIONS}
 
\vspace{1cm}
\noindent
{\bf Figure 1}
{\it Schematic description of the QCD dipole model of diffraction.} {\bf 1}-a: elastic component,
{\bf 1}-b: inelastic component

\vspace{1cm}
\noindent
{\bf Figure 2}
{\it Result of the $F_2^{D}$ fit to the H1 data \cite{F2DH194}.} 
The data are displayed by
triangles (with statistical and systematic errors added in quadrature) as a function
of $x_\Pom$ in $\beta$ and $Q^2$ bins. The fit has been performed with
statistical errors only and is displayed in full line (see text).

\vspace{1cm}
\noindent
{\bf Figure 3}
{\it Result of the $F_2^{D}$ fit to the ZEUS data \cite{zeus}.} The data are displayed by
triangles with statistical and systematic errors added in quadrature as a function
of $x_\Pom$ in $\beta$ and $Q^2$ bins. The fit has been performed with
statistical errors only and is displayed in full line (see text). For comparison, he result
of the fit of the H1 $F_2^D$ data is displayed in dashed line.

\vspace{1cm}
\noindent
{\bf Figure 4}
{\it Direct comparison between H1 (squares) and ZEUS (triangles) data.} The H1
data have been interpolated to the ZEUS bins by using the dipole parametrisation.
Both data are found compatible within error bars. Note however that sensitive 
differences are seen in particular in three different $\beta$ and $Q^2$ bins
($Q^2$=8 GeV$^2$ for $\beta$=0.2, $Q^2$=60 GeV$^2$, for $\beta$=0.7 and 0.9)
(see text).
\vspace{0.5cm}

\vspace{1cm}
\noindent
{\bf Figure 5}
{\it Effective pomeron intercept as a function of $\xpom$.} The dependence of the effective
pomeron intercept $\alpha_\Pom$ on $\xpom$ is shown for the inelastic component
(independent of $Q^2$; continuous line) and for the inelastic one
for 3 different values of $Q^2$. The value of $\beta$ is fixed at $0.3$.
The straight lines at $\alpha_\Pom$=1.395 and $\alpha_\Pom$=1.08 correspond
respectively to the bare BFKL intercept as found by the fit (see table 1),
and the soft Donnachie Landshoff prediction \cite{dola}.
\vspace{0.5cm}

\vspace{1cm}
\noindent
{\bf Figure 6}
{\it Effective pomeron intercept as a function of $Q^2$.} The effective pomeron 
intercept of the model
is displayed: elastic component 
(dashed curve), inelastic component (dotted curve), total (full curve). The 
value of $\beta$ and $\xpom$ are fixed for reference at ($\beta=0.3$
and $\xpom=10^{-3}$). The
bare and soft pomeron intercepts are given by dotted-dashed straight lines.
The H1 determination \cite{F2DH194} lies in the region between the two 
dashed straight lines. 
\vspace{0.5cm}

\vspace{1cm}
\noindent
{\bf Figure 7}
{\it Scaling violations.} The dependence of $\xpom F_2^D$ on $Q^2$ for different values of
$\beta$ at fixed $\xpom$ ($3.10^{-3}$) is shown together with 
the dipole model fit. Dotted line: elastic component, dashed line:
inelastic component, full line: total.
\vspace{0.5cm}

\vspace{1cm}
\noindent
{\bf Figure 8}
{\it Beta dependence.} The dependence of $\xpom F_2^D$ on $\beta$ for different values of
$Q^2$ at fixed $\xpom$ ($3.10^{-3}$) is shown together with 
the dipole model fit. Dotted line: elastic component, dashed line:
inelastic component, full line: total.
\vspace{0.5cm}

\vspace{1cm}
\noindent
{\bf Figure 9}
{\it Different components of the H1 $F_2^D$ fit.} Dotted line: elastic component,
dashed line: inelastic component, dashed-dotted line: secondary reggeon component,
full line: sum of all components.
\vspace{0.5cm}

\vspace{1cm}
\noindent
{\bf Figure 10}
{\it Different components of the H1 $F_2^D$ fit.} Longitudinal
versus transverse. Dotted line: tranverse elastic component,
dotted-dashed line: longitudinal elastic component, dashed line: 
total ($F_T+F_L$ with $F_L/F_T=2/9$, see text) inelastic component,
full line: result of the fit.
\vspace{0.5cm}

\vspace{1cm}
\noindent
{\bf Figure 11}
{\it Prediction for $R$, the ratio of the longitudinal to the transverse
cross section.} Note the different scale for the high $\beta$ bins.
\vspace{0.5cm}

\clearpage
\begin{center}
\psfig{figure=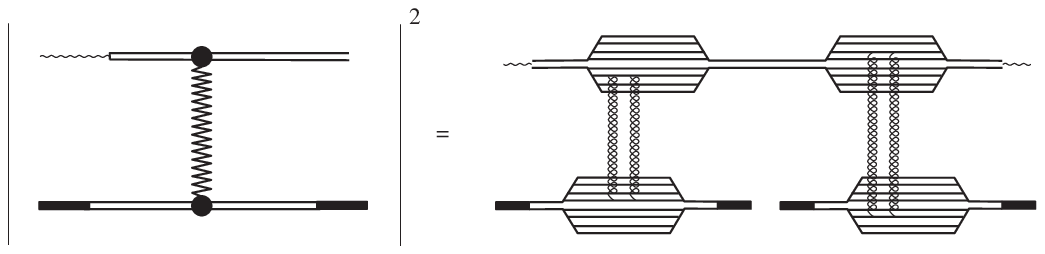,width=17cm}\\
\vskip 0.5cm
{\large\bf a.}\\
\vskip 2cm
\psfig{figure=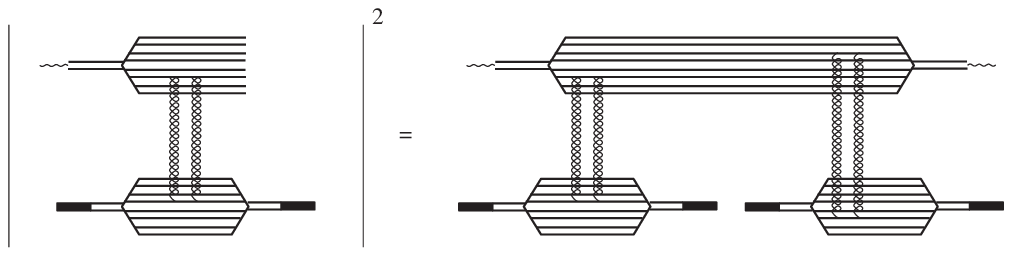,width=17cm}\\
\vskip 0.5cm
{\large\bf b.}\\
\vskip 2cm
{\large\bf Figure 1}\end{center}

\clearpage
\begin{center}
\psfig{figure=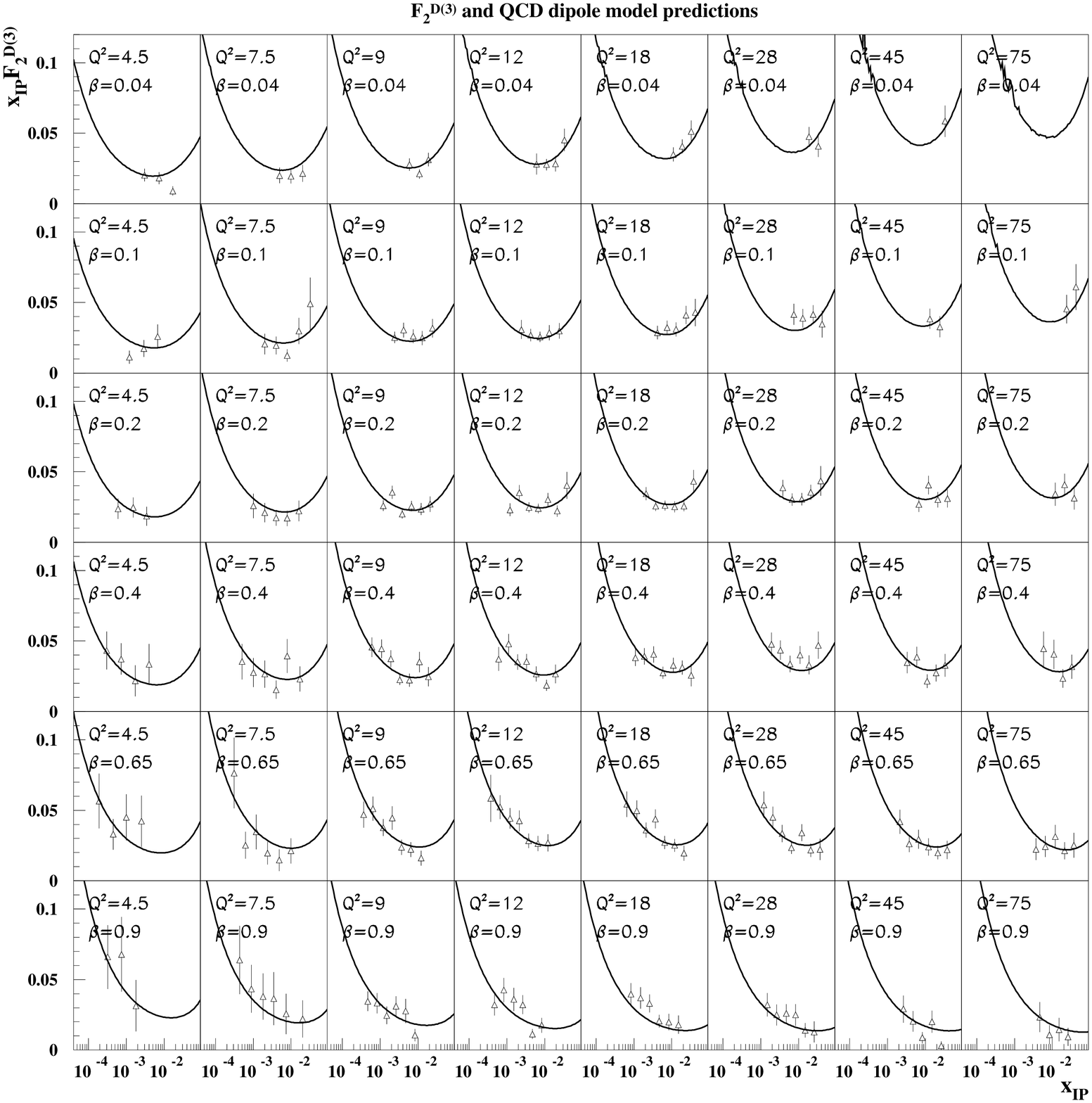,height=17cm}\\
{\large\bf Figure 2}\end{center}

\clearpage
\begin{center}
\psfig{figure=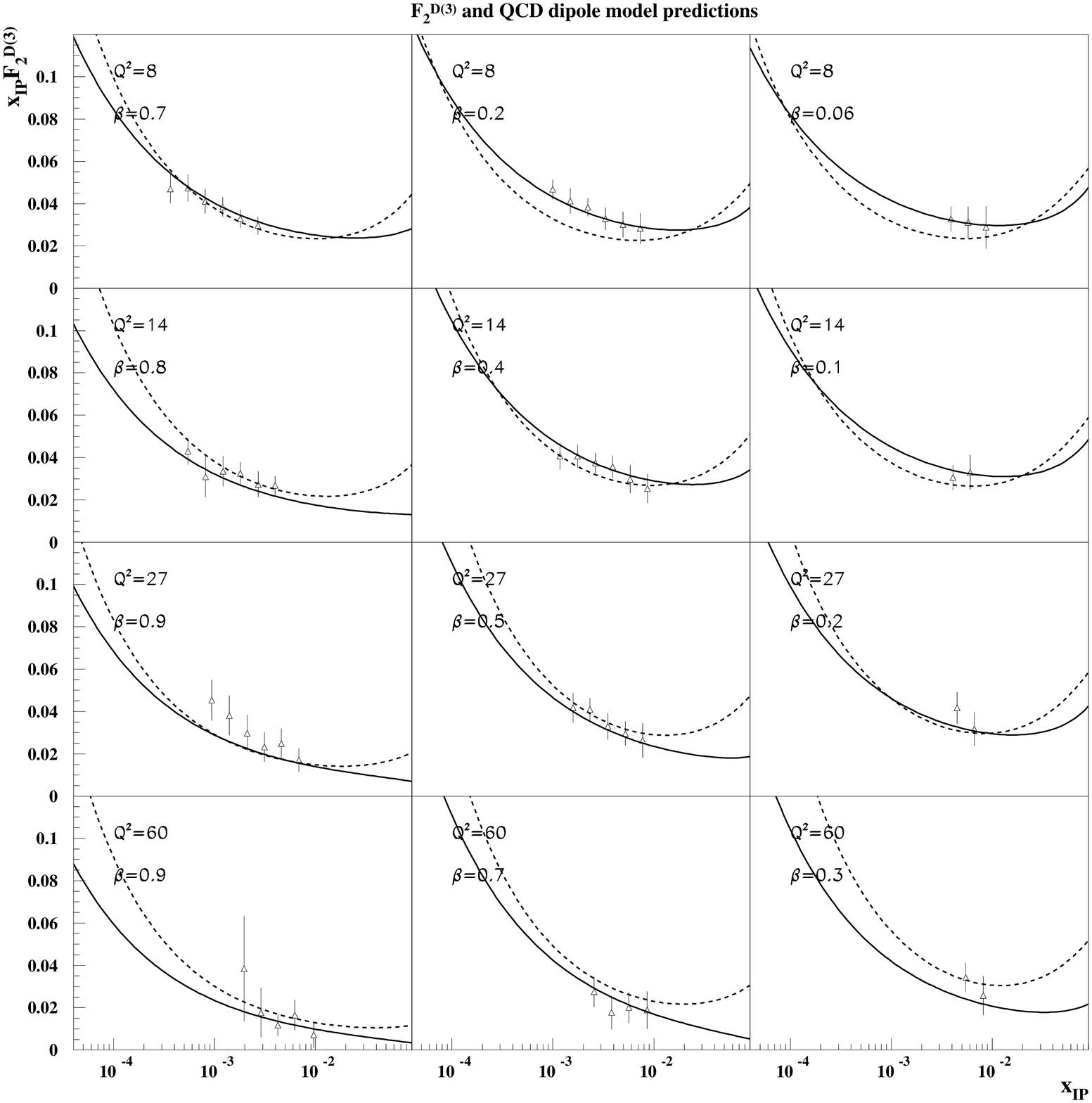,height=17cm}\\
{\large\bf Figure 3}\end{center}

\clearpage
\begin{center}
\psfig{figure=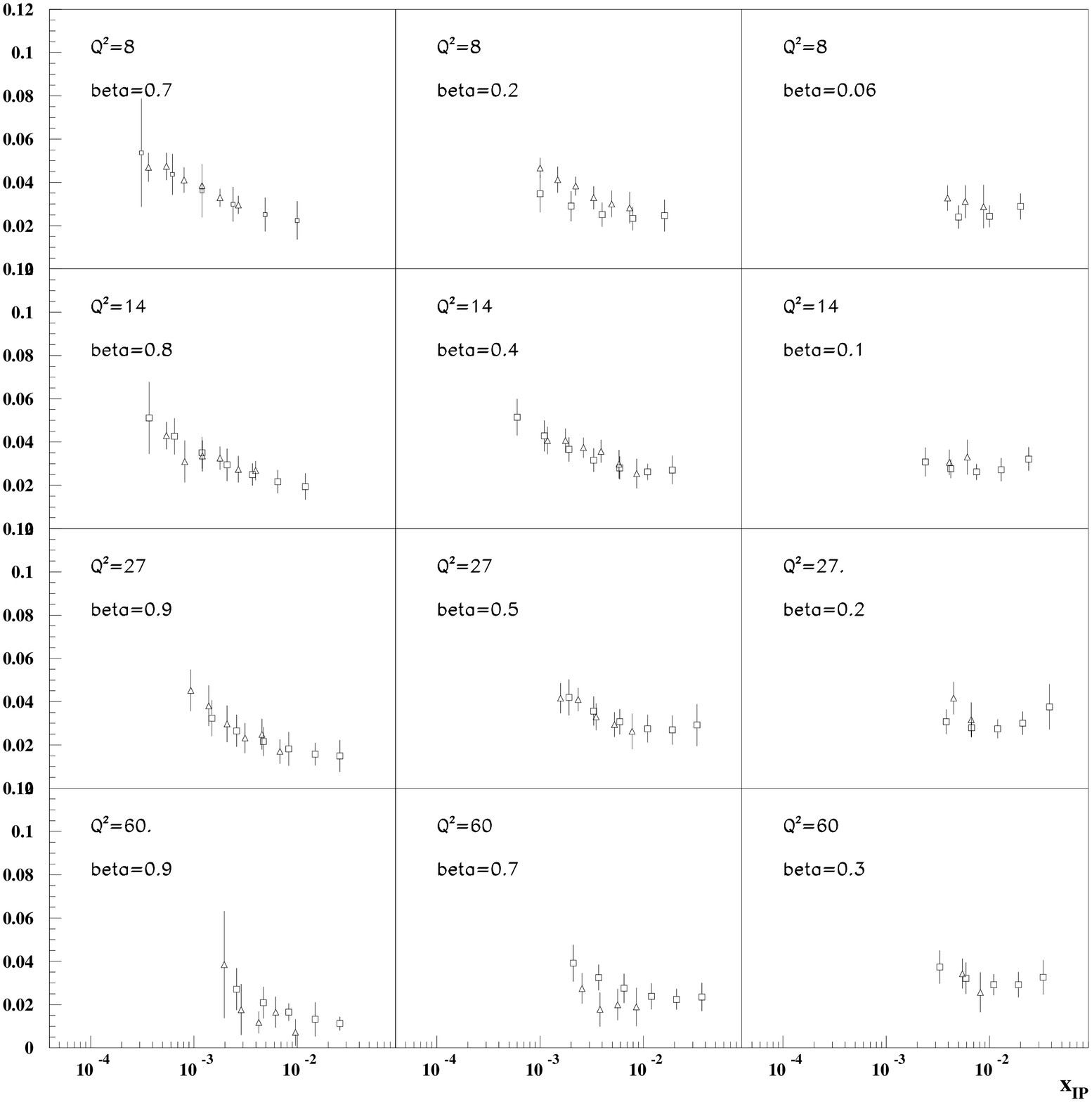,height=17cm}\\
{\large\bf Figure 4}\end{center}

\clearpage
\begin{center}
\psfig{figure=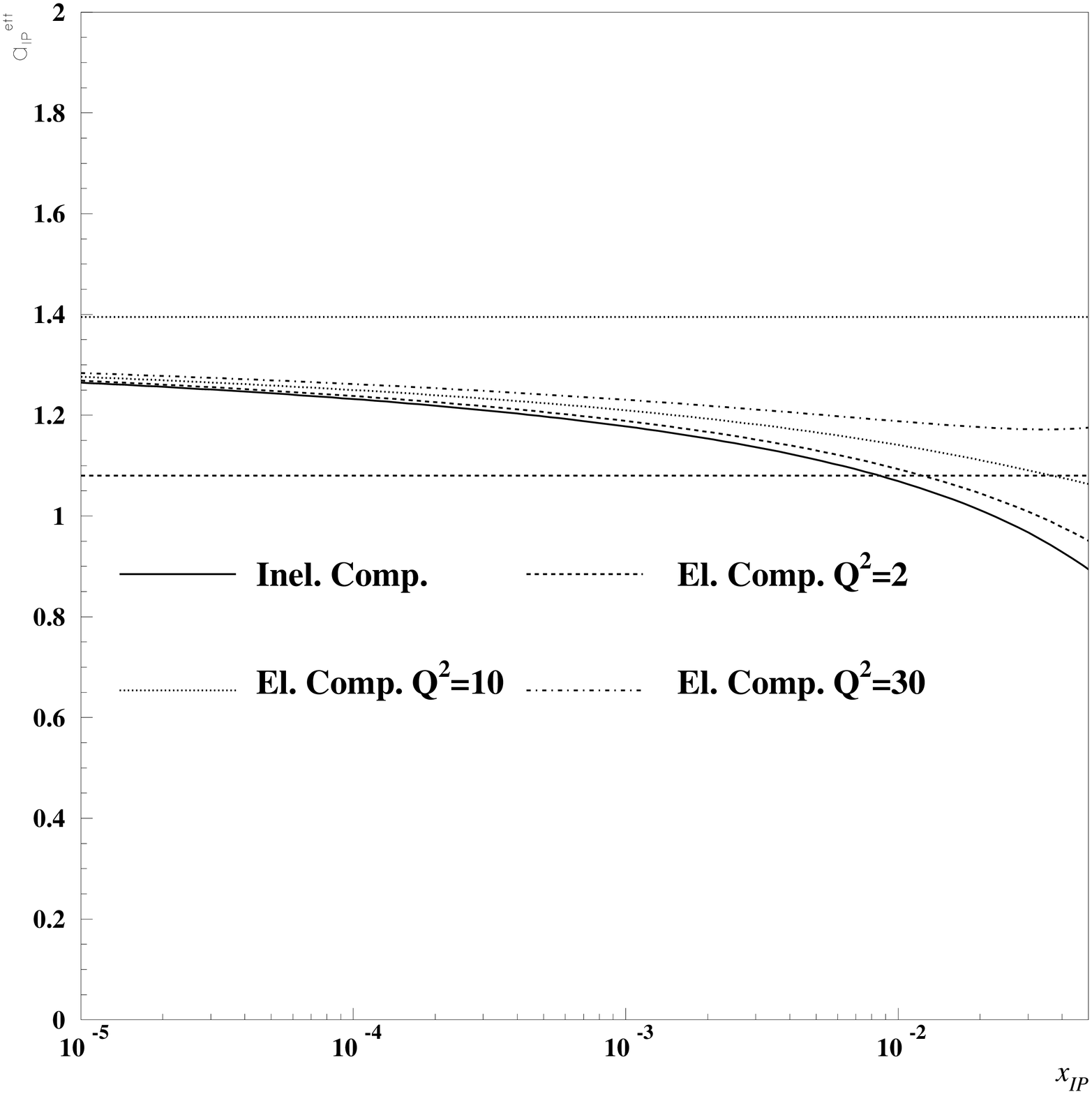,height=17cm}\\
{\large\bf Figure 5}\end{center}
\clearpage

\clearpage
\begin{center}
\psfig{figure=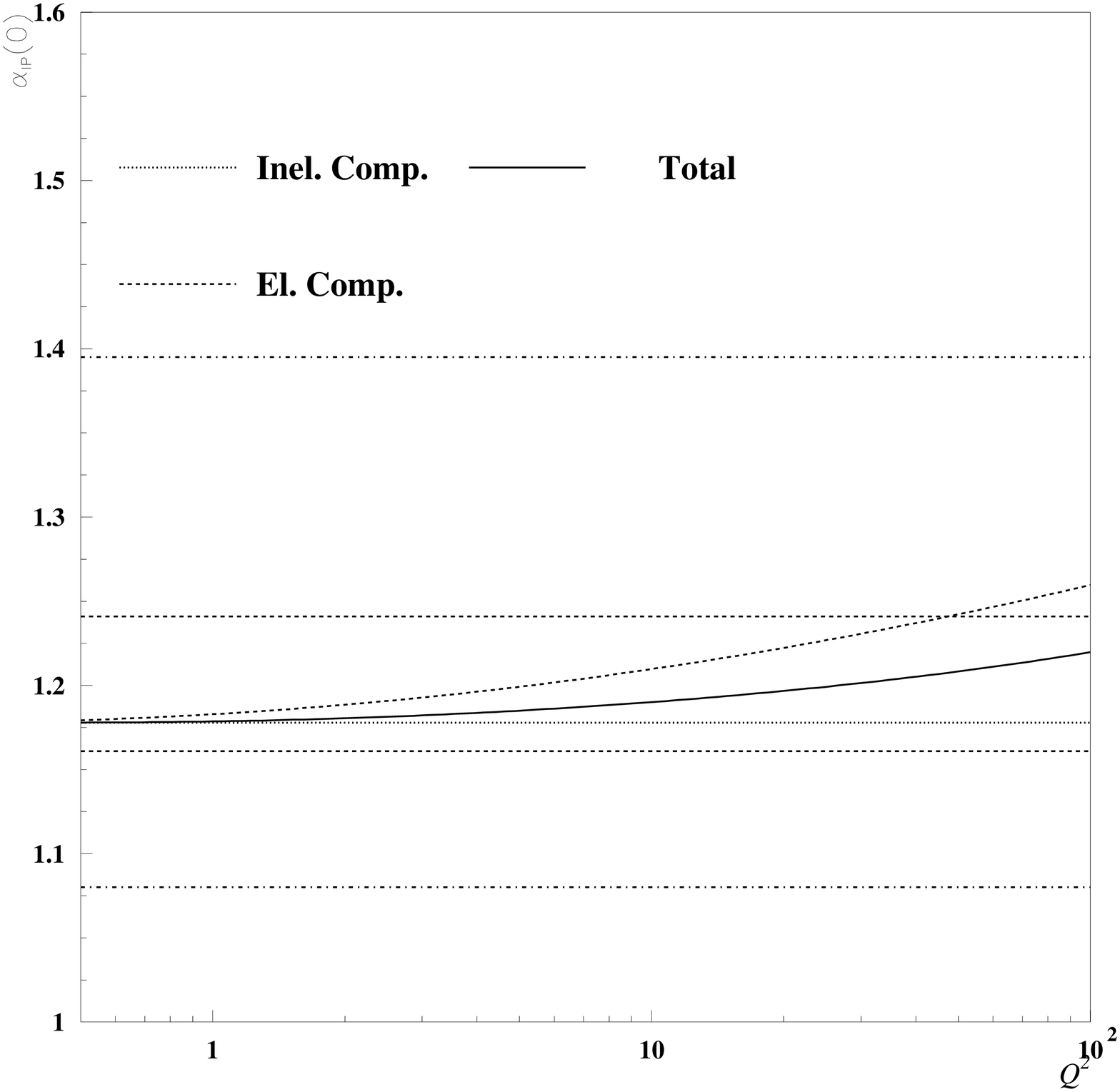,height=17cm}\\
{\large\bf Figure 6}\end{center}

\clearpage
\begin{center}
\psfig{figure=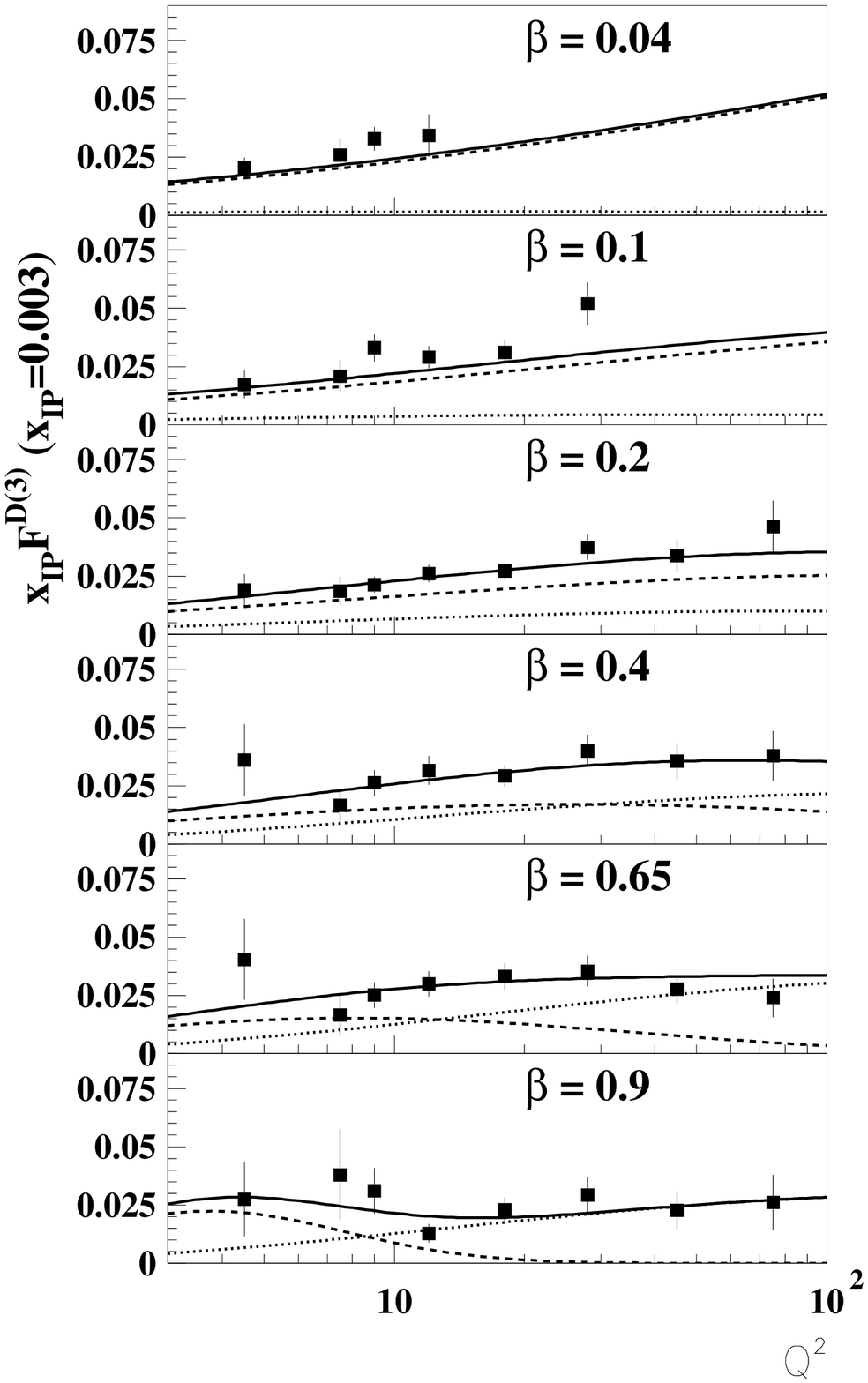,height=17cm}\\
{\large\bf Figure 7}\end{center}

\clearpage
\begin{center}
\psfig{figure=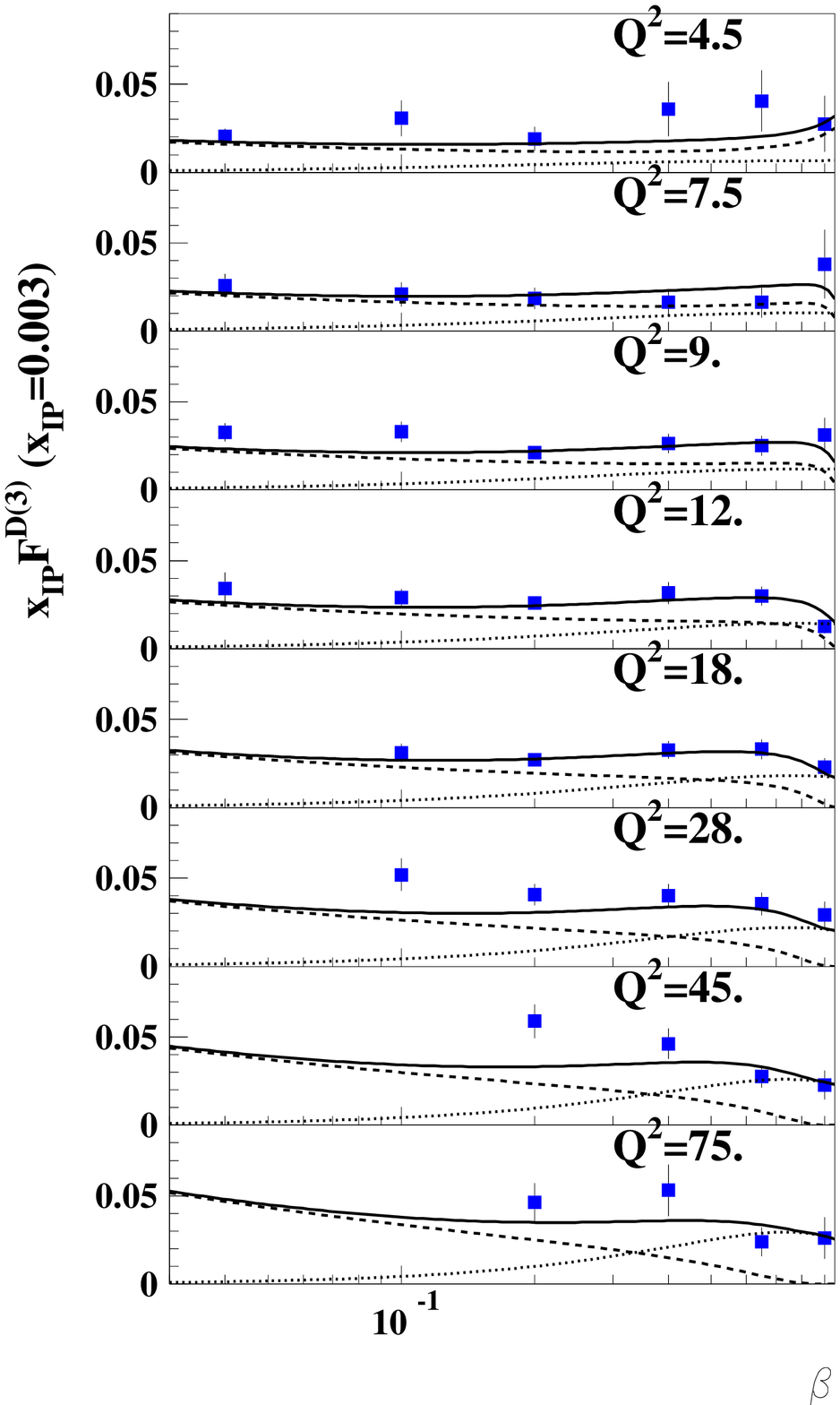,height=17cm}\\
{\large\bf Figure 8}\end{center}

\clearpage
\begin{center}
\psfig{figure=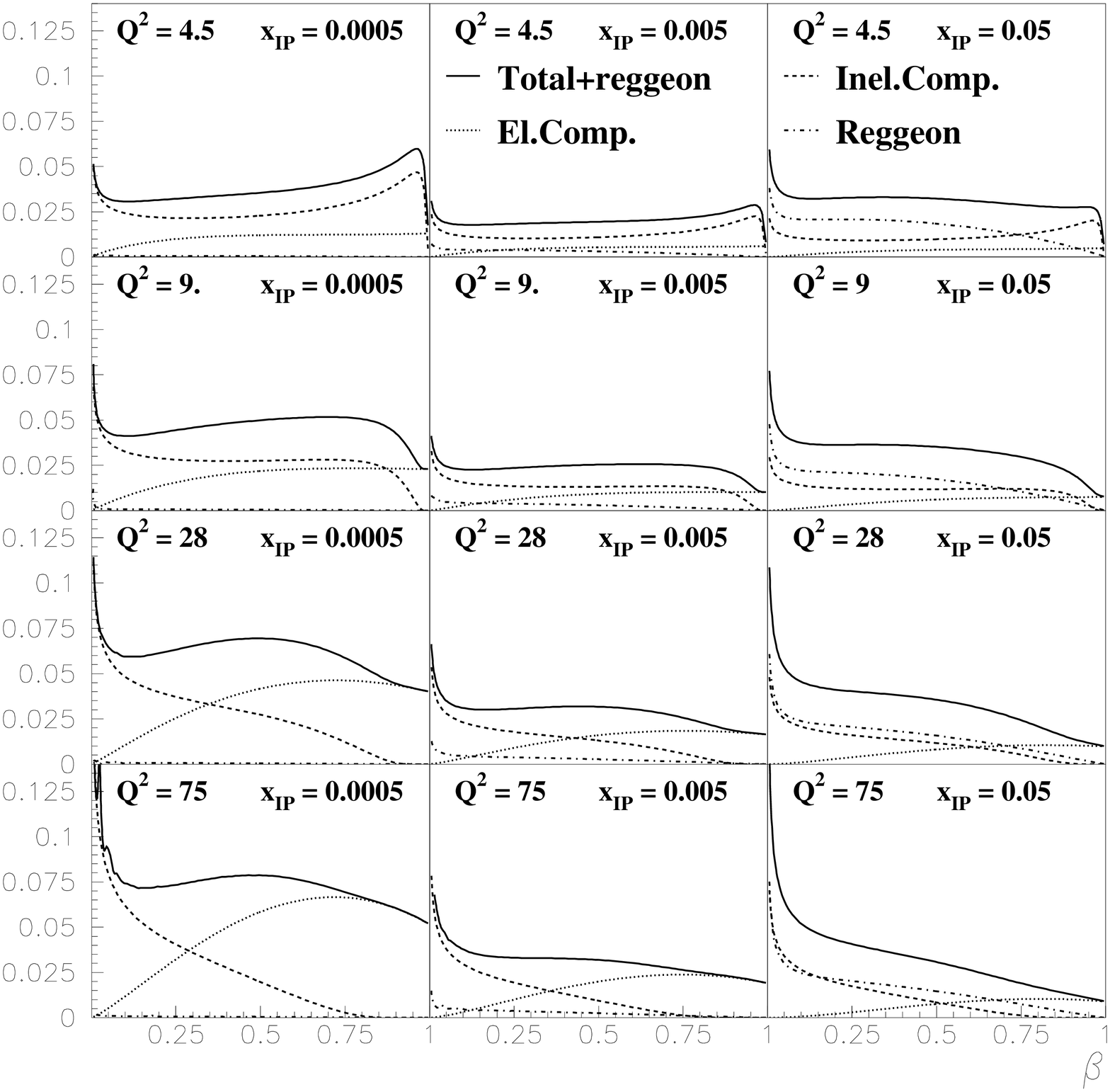,height=17cm}\\
{\large\bf Figure 9}\end{center}

\clearpage
\begin{center}
\psfig{figure=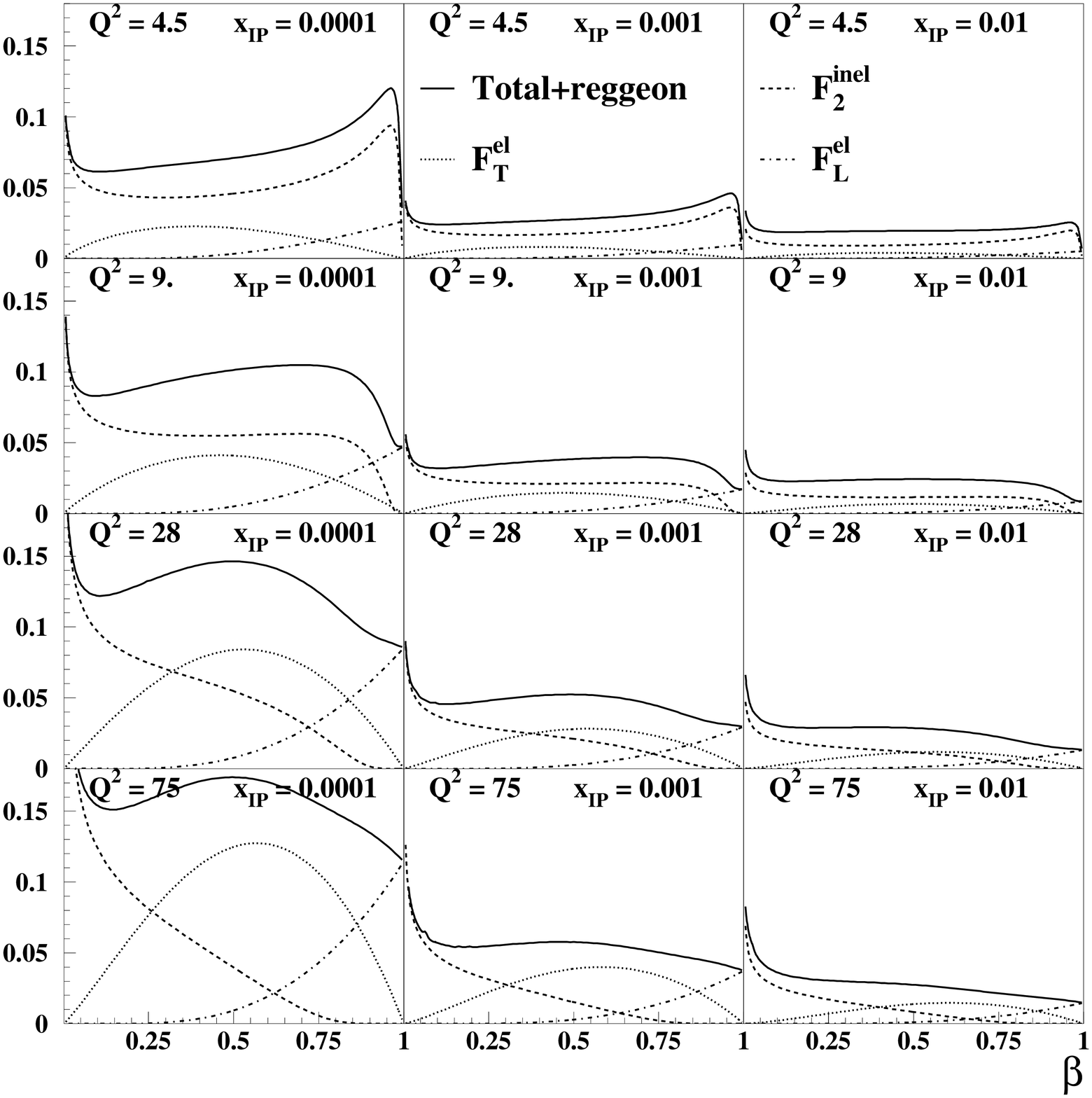,height=17cm}\\
{\large\bf Figure 10}\end{center}

\clearpage
\begin{center}
\psfig{figure=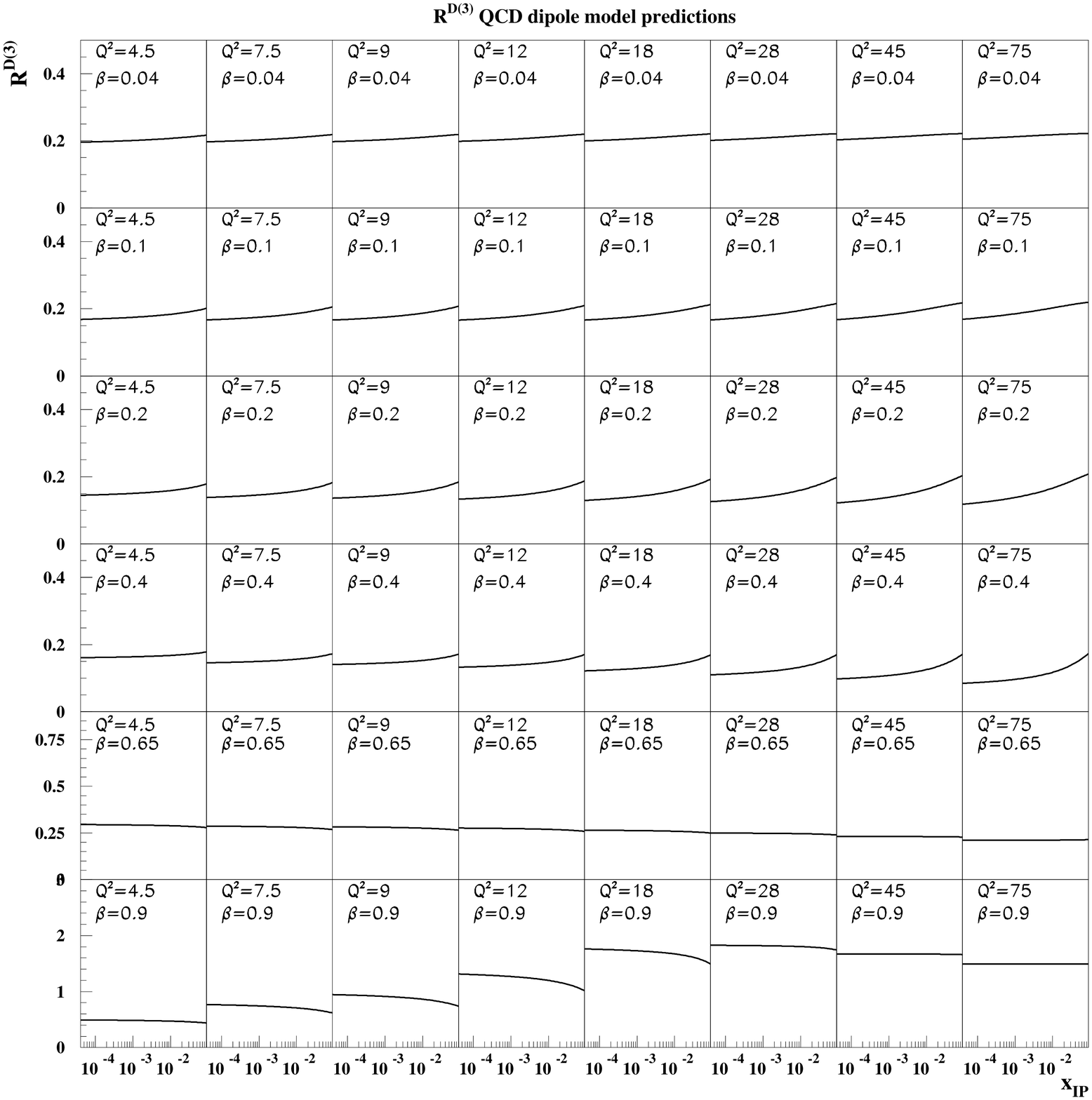,height=17cm}\\
{\large\bf Figure 11}\end{center}

\end{document}